\documentclass[%
 aip,
 jcp,
 amsmath,amssymb,
 reprint,%
]{revtex4-1}

\usepackage{graphicx}
\usepackage{dcolumn}
\usepackage{bm}

\usepackage[utf8]{inputenc}
\usepackage[T1]{fontenc}
\usepackage{mathptmx}
\usepackage{physics}
\usepackage{subfig}
\usepackage{color}

\begin{document}

\preprint{AIP/123-QED}

\title[Towards DMRG-tailored coupled cluster method in the 4c-relativistic domain]{Towards DMRG-tailored coupled cluster method\\ in the 4c-relativistic domain}

\author{Jan Brandejs}
\email{jan.brandejs@jh-inst.cas.cz}
\affiliation{J. Heyrovsk\'{y} Institute of Physical Chemistry, Academy of Sciences of the Czech \mbox{Republic, v.v.i.}, Dolej\v{s}kova 3, 18223 Prague 8, Czech Republic}
\affiliation{Faculty of Mathematics and Physics, Charles University, Prague, Czech Republic}
\author{Jakub Vi{\v s}{\v n}\'ak}
\email{jakub.visnak@jh-inst.cas.cz}
\affiliation{J. Heyrovsk\'{y} Institute of Physical Chemistry, Academy of Sciences of the Czech \mbox{Republic, v.v.i.}, Dolej\v{s}kova 3, 18223 Prague 8, Czech Republic}
\affiliation{Faculty of Mathematics and Physics, Charles University, Prague, Czech Republic}
\affiliation{Czech Academic City in Erbil, Yassin Najar street, Kurani Ankawa, Erbil, Kurdistan Region of Iraq}
\author{Libor Veis}
\email{libor.veis@jh-inst.cas.cz}
\affiliation{J. Heyrovsk\'{y} Institute of Physical Chemistry, Academy of Sciences of the Czech \mbox{Republic, v.v.i.}, Dolej\v{s}kova 3, 18223 Prague 8, Czech Republic}
\author{Maté Mihály}
\email{mate.mihaly@wigner.mta.hu}
\affiliation{Strongly Correlated Systems ``Lend{\"u}let'' Research Group,
Institute for Solid State Physics and Optics,
MTA Wigner Research Centre for Physics,
H-1121 Budapest, Konkoly-Thege Mikl{\'o}s {\'u}t 29-33, Hungary}
\affiliation{Department of Physics of Complex Systems, Eötvös Loránd University, Pf. 32, H-1518 Budapest, Hungary}
\author{\"Ors Legeza}
\email{legeza.ors@wigner.mta.hu}
\affiliation{Strongly Correlated Systems ``Lend{\"u}let'' Research Group,
Institute for Solid State Physics and Optics,
MTA Wigner Research Centre for Physics,
H-1121 Budapest, Konkoly-Thege Mikl{\'o}s {\'u}t 29-33, Hungary}
\author{Ji{\v r}{\'i} Pittner}
\email{jiri.pittner@jh-inst.cas.cz}
\affiliation{J. Heyrovsk\'{y} Institute of Physical Chemistry, Academy of Sciences of the Czech \mbox{Republic, v.v.i.}, Dolej\v{s}kova 3, 18223 Prague 8, Czech Republic}

\date{\today}

\begin{abstract}
There are three essential problems in computational relativistic chemistry: 
electrons moving at relativistic speeds, close lying states and dynamical
correlation. Currently available quantum-chemical methods are capable of solving
systems with one or two of these issues. However, there is a significant class 
of molecules, in which all the three effects are present. These are the heavier 
transition metal compounds, lanthanides and actinides with open d or f shells. 
For such systems, sufficiently accurate numerical methods are not available, 
which hinders the application of theoretical chemistry in this field.
In this paper, we combine two numerical methods in order to address this challenging class of molecules. 
These are the relativistic versions 
of coupled cluster methods and density matrix renormalization group (DMRG) method.
To the best of our knowledge, this is the first relativistic 
implementation of the coupled cluster method externally corrected by DMRG. 
The method brings a significant reduction of computational costs, as we demonstrate on the system of TlH, AsH and SbH.
\end{abstract}

\maketitle
\section{INTRODUCTION}

At the turn of the millennium, the density matrix renormalization group method (DMRG) \cite{White1992} was introduced to quantum-chemical community \cite{White1999, Chan2002, Legeza2003a} and since then, it has seen a large surge in the use for multireference systems. The biggest advantage of DMRG method is its capability to treat large active spaces, current implementations can go to about 50 active space spinors \cite{Schollwck2011, OlivaresAmaya2015}.
However, a major drawback of DMRG is its inability to capture dynamical correlation, since it cannot include all virtual spinors.  This correlation has a strong influence in the target systems of this project, which thus aims to address this problem.
The DMRG method is already well established and computational chemists started to use it, however, the methods for treating the dynamical correlations on top of DMRG are still in pioneering stage. Past efforts were either based on second order perturbation theory \cite{Kurashige2011}, internally contracted MRCI (multireference configuration interaction) \cite{Saitow2013}, random phase approximation \cite{Wouters2013}, canonical transformation method \cite{Yanai2006}, or the perturbation theory with matrix product states \cite{Ren2016}.

Our group has followed a different pathway to deal with the dynamical correlation, the coupled cluster method externally corrected by DMRG \cite{Veis2016}. As the name suggests, this is a combination of DMRG and the coupled cluster (CC) method. The CC method is known for its ability to describe dynamical correlation. In the externally corrected approach, first a DMRG calculation is performed on the strongly correlated active space, keeping the rest of the system fixed. This accounts for the static correlation. Second step is CC analysis of matrix product state (MPS) wave function, obtained from DMRG. Then a CC calculation is performed on the rest of the system, keeping in turn the active space amplitudes fixed, which captures the dynamical correlation.
Already the simplest version thereof, the tailored CCSD (CC with single and double excitations) approach \cite{Kinoshita2005,Hino2006}, yields very promising results \cite{Veis2016}. Remarkably, all previous approaches based on the use of DMRG output in another method have so far been non-relativistic, leaving the relativistic domain unexplored. This is the focus of this paper.

\begin{figure}[!ht]
    \includegraphics[width=7.7cm]{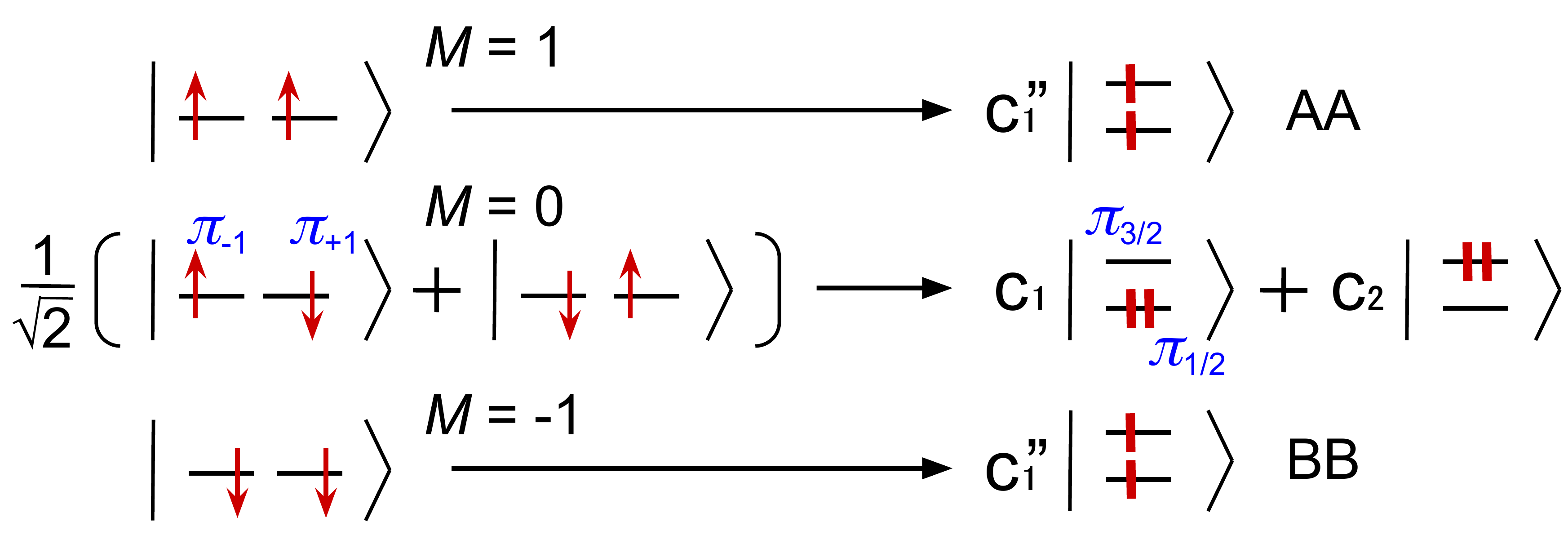}
  \caption{
  \label{fig:nitrens02}
Spin-orbit splitting in AsH and SbH.}
\end{figure}

First we demonstrate the capabilities of our relativistic 4c-TCCSD implementation on the example of the thallium hydride (TlH) molecule, which has become a standard benchmark molecule for relativistic methods and most importantly large-scale DMRG and up to CCSDTQ results are available \cite{Knecht2014}. 
It should be noted that DMRG is best suited for static-correlation problems while TlH is dominated mostly by dynamic correlation, for which CC approaches are excellent.
In order to study the behaviour of TCCSD method for more multireference systems, we performed tests on AsH and SbH molecules.
The multireference character in AsH and SbH ground state arises from the fact that two determinants are needed to describe \mbox{M$_s$ = 0} triplet component in the spin-free case, and therefore is somewhat "artificial".
Figure \ref{fig:nitrens02} depicts the $\pi_{1/2}^2$ and $\pi_{3/2}^2$ determinants arising from \mbox{$M_s$ = 0} triplet component's determinants due to spin-orbit splitting for AsH and SbH. The heavier the atom is, the greater the splitting, then $\pi_{1/2}^2$ is more dominant and X0$^+$ ground state becomes less multireference. Hence we expect AsH to be of stronger multireference nature than SbH.

\section{THEORY}
Present-day relativistic calculations are often carried out within the no-pair approximation, 
where the Dirac-Coulomb Hamiltonian is embedded by projectors eliminating the troublesome negative-energy solutions, which yields a second quantized Hamiltonian formally analogous to the non-relativistic case
\begin{eqnarray}
\label{ham2q}
H = \sum_{PQ} h_P^Q \, a_P^\dagger a_Q + \frac{1}{4}  \sum_{PQRS} \langle PQ || RS\rangle \, a^\dagger_P a^\dagger_Q a_S a_R,
\end{eqnarray}
where the indices $P,Q,R,S$ run over the positive-energy 4-component spinors spanning the one-electron basis.
The barred spinors ($\phi_{\bar{p}}$) and unbarred spinors ($\phi_{p}$) form Kramers pairs related to each other by action of the time-reversal operator $K$
\begin{eqnarray}
\label{kramers}
K \phi_p &=& \phi_{\bar{p}}, \nonumber \\ 
K \phi_{\bar{p}} &=& -\phi_p.
\end{eqnarray}

The Kramers symmetry replaces the spin symmetry in the non-relativistic theory; in particular $M_S$ is not a good quantum number
and $M_K$ projection is defined instead, which is $1/2$ for unbarred spinors (A) and $-1/2$ for spinors with barred indices (B).
The capital indices in (\ref{ham2q}) run over both spinors of a Kramers pair. 
In contrast to the non-relativistic case, the Hamiltonian (\ref{ham2q}) is in general not block-diagonal in $M_K$.
Since each creation or annihilation operator in (\ref{ham2q}) changes $M_K$ by $\pm 1/2$, the Hamiltonian couples states with $|\Delta M_K| \leq 2$.
Moreover, the index permutation symmetry of the 2e-integrals in (\ref{ham2q}) is lower than in the non-relativistic case.

The Dirac program \cite{DIRAC18} employs a quaternion symmetry approach which combines the Kramers and binary double group 
symmetry ($D_{2h}^*$ and subgroups) \cite{saue-jensen-1999}.
The double groups can be sorted into three classes based on the application of the Frobenius-Schur indicator to their 
irreducible representations: ``real groups'' ($D_{2h}^*$, $D_{2}^*$, and $C_{2v}^*$); 
``complex groups'' ($C_{2h}^*$, $C_{2}^*$, and $C_{s}^*$); and ``quaternion groups'' ($C_i^*$ and $C_1^*$) \cite{dyall-faegri}. 
Generalization of non-relativistic methods is simplest in the ``real groups'' case, where the integrals are real-valued and
the ones with odd number of barred (B) indices vanish. 
In practice, it means that additional ``spin cases'' of integrals $(\mathrm{AB}|\mathrm{AB})$ and $(\mathrm{AB}|\mathrm{BA})$ (in Mulliken notation) have to be included.
For the complex groups, the integrals are complex-valued, but still only integrals with even number of barred indices are non-zero. 
Finally, in the remaining case of  ``quaternion groups'' all the integrals have to be included 
and are complex-valued \cite{Thyssen_phd, dyall-faegri}.

The idea of externally corrected coupled cluster methods is to take information
on static correlation from some non-CC external source, and to include it into 
the subsequent CC treatment \cite{paldus-externalcorr}. 
The conceptually simplest approach is the tailored CC method (TCC)
proposed by Bartlett et al.\cite{Kinoshita2005,cyclobut-tailored-2011,melnichuk-2012,melnichuk-2014}, which uses the split-amplitude ansatz for
the wave function introduced by Piecuch et al. \cite{Piecuch1993,semi3} 
\begin{align}
\ket{\Psi} = e^{T_{\mathrm{ext}}} e^{T_{\mathrm{cas}}} \ket{ \Phi}
	\label{eqn:ansatz}
\end{align}
where $T_{\mathrm{cas}}$ containing amplitudes with all active indices is ``frozen'' at values obtained from CASCI or in our case from DMRG. 
The external cluster operator $T_{\mathrm{ext}}$ is composed of amplitudes with at least one index outside the CAS space.
Another way to justify this ansatz is the formulation of CC equations based on excitation subalgebras recently introduced by Kowalski \cite{Kowalski2018}.
The simplest version of the method truncates  both  $T_{\mathrm{cas}}$  and  $T_{\mathrm{ext}}$ to single and double excitations.
Since there is a single-determinantal Fermi vacuum, the excitation operators $T_{\mathrm{ext}}$ and $T_{\mathrm{cas}}$ commute,
which keeps the method very simple.
TCC can thus use the standard CCSD solver, modified to keep the amplitudes from $T_{\mathrm{cas}}$ fixed.
Thanks to the two-body Hamiltonian, tailored CCSD energy with the $T_{\mathrm{ext}}=0$ and $T_{\mathrm{cas}}$ from CASCI (complete active space configuration interaction) reproduces the CASCI energy. 
In the limit of CAS space including all MOs, TCC thus recovers the FCI energy.
In general, a quadratic error bound valid for TNS-TCC methods is derived \cite{Faulstich2019a}.

In \cite{Veis2016,Veis2016err} 
we have described how to obtain $T_{\mathrm{cas}}$ from the DMRG wave function using concepts of quantum information theory \cite{Legeza2003}
in the non-relativistic case, yielding the DMRG-TCCSD method.
The DMRG method
\cite{schollwock_2005} is a procedure which variationally optimizes the wave function in the form of the matrix product state (MPS) ansatz \cite{Schollwck2011}.
The quantum chemical version of DMRG (QC-DMRG) \cite{Legeza-2008,marti_2010,chan_review,wouters_review,Szalay2015,yanai_review}
eventually converges to the FCI solution in a given orbital space, i.e. to CASCI.
The practical version of DMRG is the two-site algorithm, which
provides the wave function in the two-site MPS form \cite{Schollwck2011}
\begin{eqnarray}
  \label{eq:MPS_2site}
  | \Psi_\text{MPS} \rangle = \sum_{\{\alpha\}} \mathbf{A}^{\alpha_1} \mathbf{A}^{\alpha_2} \cdots \mathbf{W}^{\alpha_i \alpha_{i+1}} \cdots \mathbf{A}^{\alpha_n}| \alpha_1 \alpha_2 \cdots \alpha_n \rangle, \nonumber \\
\end{eqnarray}
where $\alpha_i \in \{ | 0 \rangle, | \downarrow \rangle, | \uparrow \rangle, | \downarrow \uparrow \rangle \}$ and
for a given pair of adjacent indices $[i, (i+1)]$, $\mathbf{W}$ is a four index tensor, which corresponds to the eigenfunction of the electronic Hamiltonian expanded in the tensor product space of four
tensor spaces defined on an ordered orbital chain, so called \textit{left block} ($M_l$ dimensional tensor space)  , \textit{left site} (four dimensional tensor space of $i^{\text{th}}$ orbital), \textit{right site} (four dimensional tensor space of $(i+1)^{\text{th}}$ orbital), and \textit{right block} ($M_r$ dimensional tensor space).

When employing the two-site MPS wave function (Eq. \ref{eq:MPS_2site}) for the purposes of the TCCSD method,
the CI expansion coefficients $c_i^a$ and $c_{ij}^{ab}$ for $a,b,i,j \in \text{CAS}$ can be efficiently calculated by contractions of MPS matrices \cite{moritz_2007, boguslawski_2011}.
We would like to note that using the two-site DMRG approach in practice means using the wave-function calculated at different sites and it can only be employed together with the dynamical block state selection (DBSS) procedure \cite{Legeza2003a} assuring the same accuracy along the sweep. Alternatively, one can use the one-site approach in the last sweep \cite{Zgid-2008b}.

Once the CI coefficients  $c_i^a$ and $c_{ij}^{ab}$ have been obtained, the standard CC analysis is performed to convert them to the CC amplitudes
\begin{eqnarray}
        \label{eq:ci2cc}
        T^{(1)}_{\text{CAS}} & = & C^{(1)}, \\
        T^{(2)}_{\text{CAS}} & = & C^{(2)} - \frac{1}{2}[C^{(1)}]^2.
\end{eqnarray}

The generalization of the DMRG-TCCSD method  to the relativistic 4c case has to consider several points. 
First of all, the additional integral classes with nonzero $\Delta M_K$ have to be implemented in the DMRG Hamiltonian \cite{Knecht2014,Battaglia2018}.
Secondly, there will be more CI coefficients and subsequently CC amplitudes to be obtained from the MPS wave function,
corresponding to excitations with nonzero $\Delta M_K$.
Finally, except for the ``real groups'', the DMRG procedure has to work with complex matrices and the resulting cluster amplitudes will also be complex-valued.
In the present work, we have selected numerical examples with  ``real groups'' symmetry, while the complex generalization of the DMRG code is in progress.

\section{COMPUTATIONAL DETAILS}

In the present work we have used the two-site DMRG variant together
with DBSS through the course of the whole DMRG procedure and obtained the
CC amplitudes from the resulting two-site form of MPS wave function.
For all systems, DMRG calculations were performed exclusively with the QC-DMRG-Budapest program \cite{budapest_qcdmrg}. 
Dirac program package \cite{DIRAC18} was used for the remaining relativistic calculations, whereas Orca program was used for remaining non-relativistic calculations. 
Orbitals and MO integrals were generated with the Dirac program package \cite{DIRAC18}
We used the Dirac-Coulomb Hamiltonian and triple-zeta basis sets for the heavier of the two atoms (cv3z for Tl, As and Sb and cc-pVTZ for F) as well as for hydrogen (cc-pVTZ), 
which include core-correlating functions for the heavier atom. 
Initialization of DMRG, i.e., optimal ordering of spinors, was set up as discussed in Ref.~\onlinecite{Szalay2015}.
The numerical accuracy was controlled by DBSS \cite{Legeza2003a} 
keeping up to thousands of block states for the a priori set quantum information loss threshold $\chi = 10^{-6}$. 

\subsection{Comparison with non-relativistic TCC}

In order to compare with non-relativistic version of TCCSD method, the system of hydrogen fluoride was chosen,
as it is a biatomic with light nuclei. CC-pVTZ basis was used at the internuclear distance of $0.8996$~\r{A}.

Consistent methods should exhibit a constant shift of relativistic and non-relativistic 
energy $\Delta E = E_{\text{rel}} - E_{\text{nonrel}}$, given by a different Hamiltonian.
Table \ref{table:hf} shows that TCCSD is consistent with RHF, CCSD and DMRG methods in terms of $\Delta E$ up to a millihartree. 

\begin{table}[h!]
\centering
\begin{tabular}{| r | r | r | r |} 
	 \hline
 	 method & $E_{\text{rel}}$ [E$_h$] & $E_{\text{nonrel}}$ [E$_h$] & $\Delta E$ [E$_h$]\\
	  \hline
	  RHF & -100.14972 & -100.05846 & -0.09126 \\
	  DMRG(6,6) & -100.15868 & -100.06737 & -0.09130 \\
	  CCSD & -100.42322 & -100.33172 & -0.09150 \\
	  TCCSD(6,6) & -100.42418 & -100.33246 & -0.09172 \\
	   \hline
	 \end{tabular}
	\caption{
	 \label{table:hf}
		Comparison of energies of HF molecule obtained from relativistic and non-relativistic methods.
		The rightmost column shows the difference between the \mbox{4c-relativistic} energy $E_{\text{rel}}$ and non-relativistic energy $E_{\text{nonrel}}$.
	}
 \end{table}

\subsection{TlH}
\begin{figure}[!ht]
    \includegraphics[width=8.6cm]{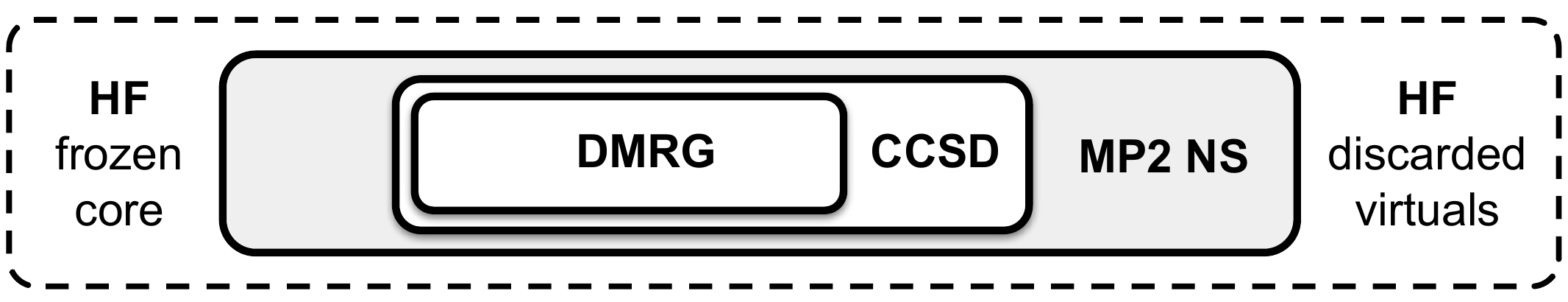}
  \caption{
  \label{fig:spaces}
  Schematic depiction of active spaces used in the 4-TCCSD procedure for TlH.}
\end{figure}

We have used the computational protocol of Ref.~\onlinecite{Knecht2014} for direct comparison with their energies. 
C$^\star_{\mathrm{2v}}$ double group symmetry with real irreps was assumed. The 4c-RHF energy was $-20275.41661$~E$_h$.
We used MP2 natural spinors (NS) from the Dirac program \cite{DIRAC18} as the spinor 
basis for electron-correlation calculations, correlating the Tl 5s, 5p, 4f 5d, 6s, 6p and H 1s electrons 
while keeping the remaining core electrons of Tl frozen. Using uncontracted basis, a virtual spinor threshold was set at 135~E$_h$.
The resulting space (14,47) was chosen by ordering MP2 NS by their occupations and taking those with values between $1.98$ and $0.001$. 
In this space, the 4c-TCCSD was performed, 
with DMRG calculations in the procedure limited to subspaces of (14,10), (14,14), (14,17), (14,25) and (14,29), with spinors sorted by MP2 occupations. 
Figure \ref{fig:spaces} shows a scheme of embedded active spaces used in the procedure.

\subsection{AsH and SbH}
Since SbH is a heavier homologue of AsH (which is itself homologue of nitrene, NH), the procedure was very similar for both of them. 
In contrast with TlH, instead of MP2 NS, we used average-of-configuration SCF spinors. The 4c-SCF energy was $-2260.04261$ ~E$_h$ for AsH and $-6481.10775$ ~E$_h$ for SbH.
The DMRG calculation in the 4c-TCCSD procedure was limited to subspaces of (16,14) and (16,23).

Dominant contributions to active spaces for AsH active space (16,14) are As: 3d,4s,4p,5s,5p, H: 1s 
and for active space (16,23) we add As: 4d, H: 2s,2p to the former. Energies of MOs are: from -2.1 E$_h$ to +0.25 E$_h$ and +1.01 E$_h$ for (16,14) and (16,23) respectively (for internuclear distance 1.52 \r{A}).
Dominant contributions to active spaces for SbH are analogous, except for principal quantum numbers, which are higher by 1.

\section{RESULTS AND DISCUSSION}
\begin{table}[h!]
\centering
\begin{tabular}{| l | r | r |} 
	 \hline
	 method & $E_{\text{el}}$ [E$_h$] & $\Delta E_{\text{el}}$ [mE$_h$] \\
	  \hline
	  4c-MP2(14,47) & -20275.85372 & -13.49 \\
	  4c-CCSD(14,47) & -20275.82966 & 10.58 \\
	  4c-CCSD(T)(14,47) & -20275.84056 & -0.32 \\
	  4c-DMRG(14,47)\footnote{4c-DMRG(14,47)[4500,1024,2048,$10^{-5}$ ], see Ref. \onlinecite{Knecht2014}.}$^,$\cite{Knecht2014} & -20275.83767 & 2.57 \\
	  4c-TCCSD(14,10) & -20275.83042 & 9.83 \\
	  4c-TCCSD(14,11) &  -20275.83170 & 8.54  \\
	  4c-TCCSD(14,12) &  -20275.83257 &  7.67 \\
	  4c-TCCSD(14,13) &  -20275.83329 &  6.95 \\
	  4c-TCCSD(14,14) & -20275.83430 & 5.94 \\
	  4c-TCCSD(14,15) & -20275.83224  &  8.00 \\
	  4c-TCCSD(14,16) & -20275.82902  & 11.23  \\
	  4c-TCCSD(14,17) & -20275.82405 & 16.19 \\
	   \hline
	 \end{tabular}
	\caption{
	 \label{table:tlh}
		Total electronic energy and energy differences $\Delta E_{\text{el}}$ (in mE$_h$) for various methods 
		with respect to the \mbox{4c-CCSDTQ(14,47)} reference energy of -20275.84024233~E$_h$ \cite{Knecht2014} for TlH  
		at the experimental equilibrium internuclear distance 1.872 \r{A}.
	}
 \end{table}

\captionsetup[subfigure]{position=b}
\begin{figure}[!ht]
  \subfloat[Comparison of TCCSD and DMRG methods. The horizontal solid line represents the ``FCI-limit'' from the large 4c-DMRG(14,47)\cite{Knecht2014} calculation. 
\label{cas_dmrg_tcc}]{%
    \includegraphics[width=8.6cm]{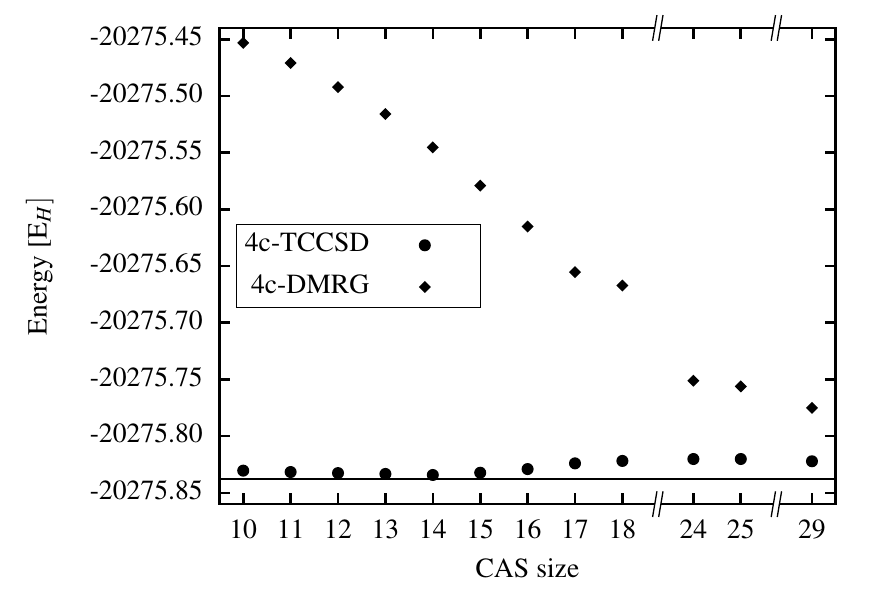}
  }
  \\
  \subfloat[Detail of 4c-TCCSD energies.\label{cassize_detail}]{%
    \includegraphics[width=8.6cm]{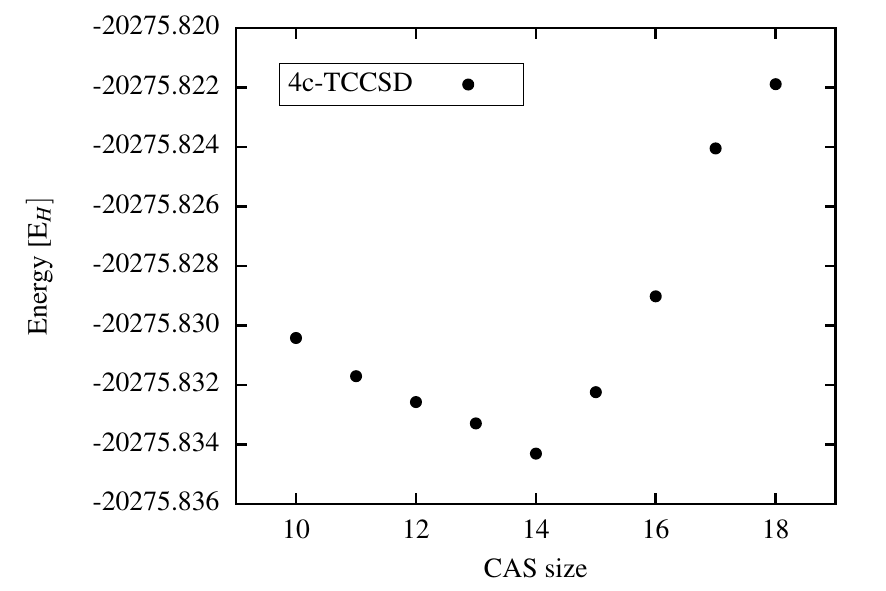}
  }
  \caption{
  \label{fig:cassize}
	Equilibrium energy of TlH calculated using the 4c-TCCSD and 4c-DMRG methods with different sizes of DMRG active space, as given in \mbox{Table \ref{table:tlh}}.}
\end{figure}

Once we reproduced the MP2 and CCSD energy of TlH in equilibrium geometry from Ref.~\onlinecite{Knecht2014}, 
we applied the \mbox{4c-TCCSD} method. Obtained energies and their respective deviations from the reference CCSDTQ calculation \cite{Knecht2014} 
are listed in Table \ref{table:tlh}. In case of the optimal selection of active space of 14-spinors, the TCCSD method improved the CCSD energy by $4.94$~mE$_H$. 
While TCCSD introduces only a minor computational cost increase over CCSD, it cuts the energy error in half.
This shows the practical advantage of the method.
The energy obtained by TCCSD is comparable even with large-scale DMRG in the full CAS(14,47).

As we can see from the high accuracy of the 4c-CCSD(T) energy, the system does not exhibit a considerable multireference character.
Therefore even a rather small CAS of 14 spinors is sufficient for a good description of the system.

As shown on the chart in Figure \ref{cas_dmrg_tcc}, TCCSD significantly improves DMRG energy towards FCI, even for the smallest CAS space.
In fact, further enlarging of CAS over the size of 14 spinors is counter productive.
Although the TCC must reproduce FCI energy when CAS is extended to all spinors, the TCC energy does not approach 
this limit monotonically \cite{Faulstich2019b}.
The obvious reason is that the ``frozen'' $T_{\text{CAS}}$ amplitudes cannot reflect the influence of the dynamical correlation 
in the external space back on the active CAS space, therefore extending CAS space first exacerbates the results. 
Unfortunately, more detailed understanding of this highly nonlinear behavior ia still an open problem.
In spite of a considerable effort, so far no quantities able to predict the optimal CAS-EXT split a priori have been identified \cite{Faulstich2019b}. 
Nevertheless, we have found that an error minimum can be obtained by sweeping through the
entire orbital space with low cost DMRG calculations and this minimum does not
shift or shifts only a little when more accurate calculations are performed.
Therefore, in practice, the optimal CAS size related to the energy minimum is
usually independent of M and can be determined with a low bond dimension (M)
DMRG calculations\cite{Faulstich2019b}.

As demonstrated by the chart \ref{cassize_detail}, 
the optimal CAS size is 14 spinors for the equilibrium energy calculation. 
This CAS size is optimal not only for energies, but also for the calculation of spectroscopic properties, including the low bond dimension calculations with $M$=512 (see \mbox{Table \ref{table:spec}}).

Table \ref{table:spec} shows the obtained spectroscopic properties of TlH. Even for a small active space of 10 spinors,
TCCSD shows an agreement with the experiment comparable with the large DMRG(14,47) calculation. For the 14-spinor space, spectroscopic constants obtained by TCCSD exhibit the best agreement with the experiment, thus being consistent with the lowest energy single point result of 14-spinor space in \mbox{Table \ref{table:tlh}}.
Moreover,  TCCSD based on DMRG with  M=512 states or on DMRG with DBSS
	yield very similar results, indicating that the underlying DMRG is well
	converged.
However, for 17-spinors, there is a bigger difference and M=512 might not be accurate enough.
This is in accordance with the previous findings of dynamical correlation effects. 

\begin{table}[h!]
\centering
\begin{tabular}{| l | r | r | r |} 
	 \hline
	 method & $r_e$ [{\r A}] & $\omega_e$ [cm$^{-1}$] & $\omega_ex_e$ [cm$^{-1}$]\\
	  \hline
	  experiment\footnote[1]{GRECP spin–orbit MRD-CI, see Ref. \onlinecite{Titov2001}.}$^,$\cite{Titov2001} & 1.872 & 1391 & 22.7 \\
	  4c-DMRG(14,47) \cite{Knecht2014} & 1.873 & 1411 & 26.6 \\
	  4c-CCSD(14,47) \cite{Knecht2014} & 1.871 & 1405 & 19.4 \\
		4c-TCCSD(14,10)\footnote[2]{TWOFIT 4th order polynomial.} DBSS & 1.874	 & 1404	 & 24.6  \\
	  4c-TCCSD(14,10)\footnotemark[2] M=512 & 1.874 & 1403 & 23.4 \\
	  4c-TCCSD(14,14)\footnotemark[2] DBSS & 1.869 &	1412 &	22.6  \\
	  4c-TCCSD(14,14)\footnote[3]{VIBANAL 10th order polynomial. R$_\text{min}$-R$_\text{max}$ [\r{A}]: 1.64-2.20 for (14,14); 1.70-2.04 for (14,17) DBSS and 1.72-2.00 for (14,17) M=512.} DBSS & 1.869 & 1411 & 20.1 \\
	  4c-TCCSD(14,14)\footnotemark[2] M=512 & 1.869 &	1411 &	22.6 \\
	  4c-TCCSD(14,14)\footnotemark[3] M=512 & 1.869 & 1411 & 20.3 \\
	  4c-TCCSD(14,17)\footnotemark[2] DBSS & 1.859 &	1426 &	20.5   \\
	  4c-TCCSD(14,17)\footnotemark[3] DBSS & 1.859  & 1426 & 17.5 \\
	  4c-TCCSD(14,17)\footnotemark[2] M=512 & 1.859 &	1428 &	22.4  \\
	  4c-TCCSD(14,17)\footnotemark[3] M=512 & 1.859 & 1428 & 29.8 \\
	   \hline
	 \end{tabular}
	\caption{
	 \label{table:spec}
	 Spectroscopic constants of ${}^{205}$TlH obtained from 4c-TCCSD, compared with calculations and experimental work from the literature.	
 The spectroscopic constants have been evaluated from potential energy curve fit, with two different methodologies. In case of TWOFIT methodology, the number of points have been selected according to Mean displacement in harmonic ground state criterion. In case of VIBANAL methodology, a wider symmetric interval around equilibrium geometry has been selected. In all cases, internuclear separation axis sampling was chosen to be $0.02$~\r{A}.}
 \end{table}
\begin{figure}[!ht]
\vskip -0.2cm
    \includegraphics[width=8.6cm]{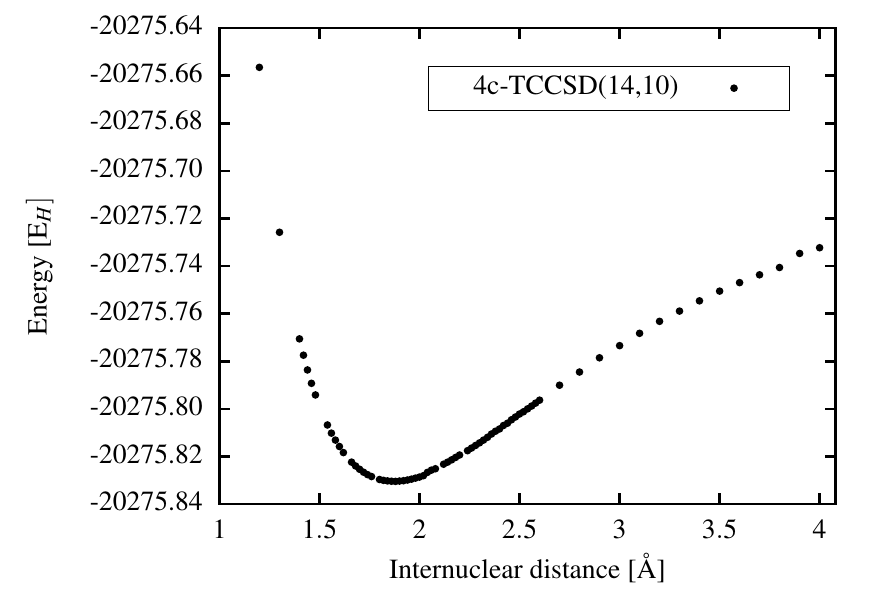}
  \caption{
  \label{fig:curve}
  Dissociation curve of TlH.}
\end{figure}

In order to further assess the feasibility of the method for multireference systems, we studied AsH and SbH molecules. Table \ref{table:conf} compares the determinants with highest coefficients as calculated by 4c-DMRG with the space of (14,14) for TlH and (16,14) for AsH and SbH. The coefficients confirm that the need arises for a multireference description of the ground state of AsH and that SbH is between AsH and TlH in terms of its multireference character.

\begin{figure}[!ht]
\vskip -0.2cm
    \includegraphics[width=8.6cm]{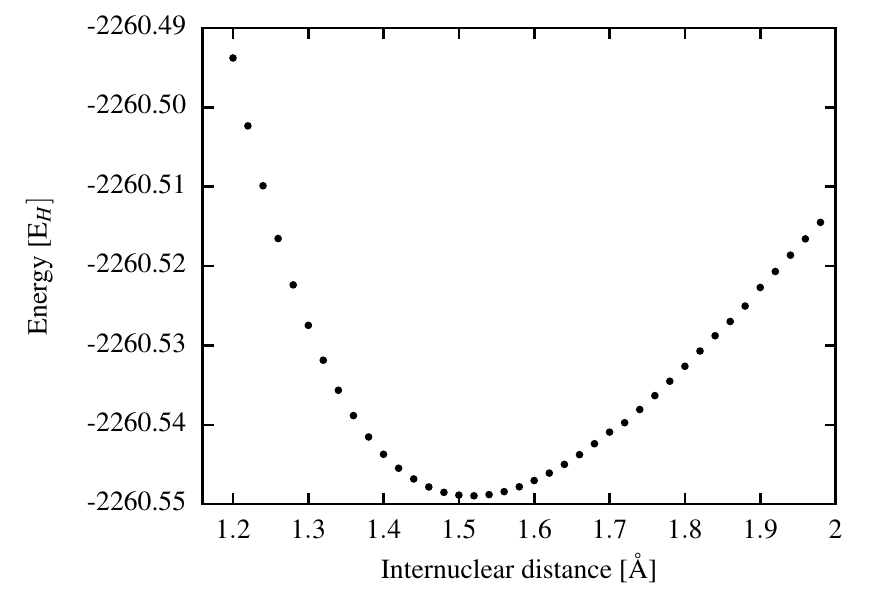}
  \caption{
  	\label{fig:ash_curve}
  Detail of potential curve of AsH near the equilibrium internuclear distance.
}
\end{figure}

The equilibrium energies of AsH in Table V show that 4c-TCCSD improved the 4c-CCSD by $17~$mE$_h$, with just $3~$mE$_h$ difference from 4c-CCSD(T). In this case a more accurate theoretical reference energy is unavailable, and therefore we turned to spectroscopic constants, as shown in Table \ref{table:ash_spec}, to enable a comparison with accurate IR spectra obtained by CO laser magnetic resonance in Ref. \onlinecite{Hensel1995}. The respective potential curve is plotted in Figure \ref{fig:ash_curve}. The comparison of both the internuclear distance $r_e$ and the vibrational constant $\omega_e$ shows a clear advantage of 4c-TCCSD over 4c-CCSD in this case, which we attribute to the multireference nature of AsH.
	Despite the multireference character, the 4c-CCSD(T) with perturbative triples still prevails over 4c-TCCSD.
This shows that at this range of internuclear distances the multireference
character is not strong enough to cause CCSD(T) to fail. Nevertheless,
in AsH it
is strong enough that TCCSD provides a major improvement over CCSD at the
same computational scaling.

\begin{table}[h!]
\centering
\begin{tabular}{| l | c | r | } 
	 \hline
	 & determinant & coefficient  \\
	  \hline
	  \hline
TlH & 2 2 2 2 2 2 2 0 0 0 0 0 0 0 & reference \\
	  \hline
    & 2 2 2 2 2 2 2 0 0 0 0 0 0 0     &     0.97862 \\
		& 2 2 2 2 2 2 0 2 0 0 0 0 0 0     &     0.05928  \\
		& 2 2 2 2 2 2 0 0 2 0 0 0 0 0     &     0.05540  \\
	   \hline
	   \hline
SbH & 2 2 2 2 2 2 2 2 0 0 0 0 0 0 & reference \\
	  \hline
    & 2 2 2 2 2 2 2 2 0 0 0 0 0 0     &     0.82518 \\
	  & 2 2 2 2 2 2 2 0 2 0 0 0 0 0     &     0.54748 \\
	  & 2 2 2 2 2 2 0 2 0 2 0 0 0 0     &     0.04872 \\
	   \hline
	   \hline
AsH & 2 2 2 2 2 2 2 2 0 0 0 0 0 0 & reference \\
	  \hline
    & 2 2 2 2 2 2 2 2 0 0 0 0 0 0    &      0.75673 \\
	  & 2 2 2 2 2 2 2 0 2 0 0 0 0 0    &      0.64048 \\
	  & 2 2 2 2 2 2 0 2 0 2 0 0 0 0    &      0.04123 \\
	   \hline
	 \end{tabular}
	\caption{
	 \label{table:conf}
 	 Three configurations with the highest coefficients (in absolute values) for TlH, SbH and AsH in equilibrium internuclear distance, as generated by DMRG with the active space of 14 spinors. In each case, the CC amplitudes were generated with respect to the closed shell reference listed in the first row for given system. Here 2 is for a doubly occupied Kramers pair and 0 is for an empty Kramers pair.
	}
 \end{table}

\begin{table}[h!]
\centering
\begin{tabular}{| l | r | } 
	 \hline
	 method & $E_{\text{el}}$ [E$_h$]  \\
	  \hline
	  4c-MP2(16,81) & -2260.53227  \\
	  4c-CCSD(16,81) & -2260.53241  \\
	  4c-TCCSD(16,14) & -2260.54945  \\
	  4c-CCSD(T)(16,81) & -2260.55133  \\
	   \hline
	 \end{tabular}
	\caption{
	 \label{table:ash}
	 Total energy for various methods for AsH  
	 at the experimental equilibrium internuclear distance\cite{Balasubramanian1989} of 1.5343~\r{A}.
	}
 \end{table}

\begin{table}[h!]
\centering
\begin{tabular}{| l | r | r | r |} 
	 \hline
	 method & $r_e$ [{\r A}] & $\omega_e$ [cm$^{-1}$] & $\omega_ex_e$ [cm$^{-1}$]\\
	  \hline
	  experiment\cite{Hensel1995} & 1.523 & 2156 & 39.2 \\
	  4c-SCF\footnote[5]{TWOFIT 4th order polynomial.} & 1.513 & 2382 & 33.6 \\
	  4c-MP2(16,81)\footnotemark[5] & 1.503 &	2256 &	35.0 \\
	  4c-DMRG(16,14)\footnotemark[5] DBSS & 1.545  &	2051 &	42.2  \\
	  4c-CCSD(16,81)\footnotemark[5] & 1.505 &	2281 &	40.0 \\
	  4c-TCCSD(16,14) \footnote[4]{VIBANAL 10th order polynomial.} DBSS & 1.517 &	2154 &	43.9  \\
	  4c-TCCSD(16,14)\footnotemark[5] DBSS & 1.517 &	2172 &	39.6  \\
	  4c-CCSD(T)(16,81)\footnotemark[5] & 1.521 & 	2146	&  40.6 \\
	   \hline
	 \end{tabular}
	\caption{
	 \label{table:ash_spec}
 	 Spectroscopic constants of $^{75}$AsH obtained from 4c-TCCSD, compared with calculations and experimental work from the literature.
 }
 \end{table}

Calculated equilibrium energies for SbH are listed in Table \ref{table:sbh}.
Compared with the higher order methods, the 4c-MP2 method outputs lower energy,
which might be due to a failure to describe the multireference character of this system. 
As with the previous systems, TCCSD energy is between CCSD and CCSD(T). 
However, in constrast with TlH where static correlation was not essential for a good description,
he it is not clear if CCSD(T) is a good benchmark, since for multireference systems CCSD(T) tends to output too low energy.
Unfortunately, we miss a more accurate calculation to compare with, hence we again turn to spectroscopic constants,
which are listed in Table \ref{table:sbh_spec}.
Considering the calculated internuclear distance, MP2 and DMRG are inaccurate, as they were for AsH. 
Oddly enough, plain DMRG with a small active space is in both cases very accurate for $\omega_e$ and $\omega_ex_e$.
Methods from CC theory succeeded in describing $r_e$, $\omega_e$ and $\omega_ex_e$, and TCCSD again outputs values between CCSD and CCSD(T).
Compared with CCSD, TCCSD improved the vibrational constant $\omega_e$.

\begin{table}[h!]
\centering
\begin{tabular}{| l | r | } 
	 \hline
	 method & $E_{\text{el}}$ [E$_h$]  \\
	  \hline
	  4c-MP2(16,81) & -6481.69651  \\
	  4c-CCSD(16,81) & -6481.67108  \\
	  4c-TCCSD(16,14) & -6481.68380  \\
	  4c-CCSD(T)(16,81) & -6481.69315  \\
	   \hline
	 \end{tabular}
	\caption{
	 \label{table:sbh}
	 Total energy for various methods for SbH  
	 at the equilibrium internuclear distance\cite{Beutel1996} of 1.70187~\r{A}.
	}
 \end{table}

\begin{table}[h!]
\centering
\begin{tabular}{| l | r | r | r |} 
	 \hline
	 method & $r_e$ [{\r A}] & $\omega_e$ [cm$^{-1}$] & $\omega_ex_e$ [cm$^{-1}$]\\
	  \hline
	  experiment\cite{Beutel1996} & 1.702 & & \\
	  experiment\cite{Hensel1995} & 1.711 & 1897 &  \\
	  experiment\cite{Urban1993} &  & 1923 & 34.2 \\
	  4c-SCF\footnote[6]{TWOFIT 4th order polynomial.} & 1.705 &	2024  &	28.0 \\
	  4c-SCF\footnote[7]{VIBANAL 10th order polynomial.} & 1.704 &	2043 &	38.1 \\
	  4c-MP2(16,81)\footnotemark[6] & 1.693 &	2009 &	30.2 \\
	  4c-DMRG(16,14)\footnotemark[6] M=2200 &  1.737 &	1839 &	35.3 \\
	  4c-CCSD(16,81)\footnotemark[6] & 1.706 &	1945 &	32.6 \\
	  4c-TCCSD(16,14)\footnotemark[6] M=2200 & 1.706 &	1937 &	36.4 \\
	  4c-CCSD(T)(16,81)\footnotemark[6] & 1.710 &	1916 &	35.3 \\
	   \hline
	 \end{tabular}
	\caption{
	 \label{table:sbh_spec}
 	 Spectroscopic constants of $^{121}$SbH obtained from 4c-TCCSD, compared with calculations and experimental work from the literature.
 }
 \end{table}

\begin{figure}[!ht]
\vskip -0.2cm
    \includegraphics[width=8.6cm]{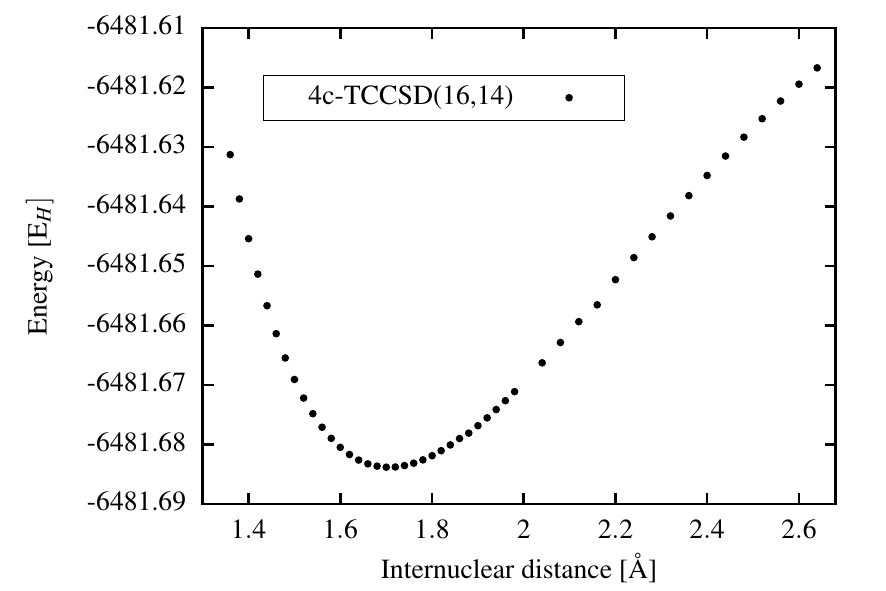}
  \caption{
  	\label{fig:sbh_curve}
  Detail of potential curve of SbH near the equilibrium internuclear distance.
}
\end{figure}

\section{CONCLUSION}
We have implemented the relativistic tailored coupled clusters method, 
which is capable of treating relativistic, strongly correlated 
systems both in terms of static and dynamical correlation.
The aim was to show that compared with the previously published calculations, we can obtain results of equal quality with much smaller active space, i.e. at a fraction of computational cost.
The results presented are promising. 
Even with a small active space, the new method 
showed comparable performance for TlH to DMRG with large CAS(14,47).
The optimal CAS size related to the energy minimum  
was determined with low cost DMRG calculations.
Calculated spectroscopic properties of TlH agree with experimental values within the error bounds.
	The comparison with experimental spectroscopic constants for AsH,
	which has stronger multireference character, has shown that TCCSD is able
	to describe such systems more accurately that CCSD, with a computational
	cost lower than CCSD(T).

\begin{acknowledgments}
	We thank developers of the Dirac program, in particular Dr. Lucas Visscher and Dr. Stefan Knecht, for providing
	access to the development version of the code and for helpful discussions.
	The work of the Czech team has been supported by the \textit{Czech Science Foundation} Grant \mbox{No. 18-24563S}.
	\mbox{{\"O}. Legeza} has been supported by the Hungarian National Research, Development and Innovation Office (NKFIH) through Grant \mbox{No. K120569},  
  the Hungarian Quantum Technology National Excellence Program, project No. \mbox{2017-1.2.1-NKP-2017-00001}. 
  The development of the relativistic DMRG
  libraries was supported by the Center for Scalable and Predictive methods
  for Excitation and Correlated phenomena (SPEC), which is funded as part of
  the Computational Chemical Sciences Program by the U.S. Department of
  Energy (DOE), Office of Science, Office of Basic Energy Sciences, Division of
  Chemical Sciences, Geosciences, and Biosciences at Pacific Northwest National Laboratory.
	\mbox{M. Mihály} has been supported by the ÚNKP-19-3 New National Excellence Program of the Ministry for
	Innovation and Technology.
  Mutual visits with the Hungarian group have been partly
	supported by the Hungarian-Czech Joint Research Project MTA/19/04.
	Part of the CPU time for the numerical computations has been supported by The Ministry of Education, 
	Youth and Sports from the Large Infrastructures for Research, Experimental Development and 
	Innovations project ``IT4Innovations National Supercomputing Center – LM2015070''. Access to computing and storage facilities owned by parties and projects contributing to the National Grid Infrastructure MetaCentrum provided under the programme ``Projects of Large Research, Development, and Innovations Infrastructures'' (CESNET LM2015042), is appreciated.
\end{acknowledgments}

\nocite{*}
\bibliography{reltcc}

\begin{thebibliography}{47}%
\makeatletter
\providecommand \@ifxundefined [1]{%
 \@ifx{#1\undefined}
}%
\providecommand \@ifnum [1]{%
 \ifnum #1\expandafter \@firstoftwo
 \else \expandafter \@secondoftwo
 \fi
}%
\providecommand \@ifx [1]{%
 \ifx #1\expandafter \@firstoftwo
 \else \expandafter \@secondoftwo
 \fi
}%
\providecommand \natexlab [1]{#1}%
\providecommand \enquote  [1]{``#1''}%
\providecommand \bibnamefont  [1]{#1}%
\providecommand \bibfnamefont [1]{#1}%
\providecommand \citenamefont [1]{#1}%
\providecommand \href@noop [0]{\@secondoftwo}%
\providecommand \href [0]{\begingroup \@sanitize@url \@href}%
\providecommand \@href[1]{\@@startlink{#1}\@@href}%
\providecommand \@@href[1]{\endgroup#1\@@endlink}%
\providecommand \@sanitize@url [0]{\catcode `\\12\catcode `\$12\catcode
  `\&12\catcode `\#12\catcode `\^12\catcode `\_12\catcode `\%12\relax}%
\providecommand \@@startlink[1]{}%
\providecommand \@@endlink[0]{}%
\providecommand \url  [0]{\begingroup\@sanitize@url \@url }%
\providecommand \@url [1]{\endgroup\@href {#1}{\urlprefix }}%
\providecommand \urlprefix  [0]{URL }%
\providecommand \Eprint [0]{\href }%
\providecommand \doibase [0]{http://dx.doi.org/}%
\providecommand \selectlanguage [0]{\@gobble}%
\providecommand \bibinfo  [0]{\@secondoftwo}%
\providecommand \bibfield  [0]{\@secondoftwo}%
\providecommand \translation [1]{[#1]}%
\providecommand \BibitemOpen [0]{}%
\providecommand \bibitemStop [0]{}%
\providecommand \bibitemNoStop [0]{.\EOS\space}%
\providecommand \EOS [0]{\spacefactor3000\relax}%
\providecommand \BibitemShut  [1]{\csname bibitem#1\endcsname}%
\let\auto@bib@innerbib\@empty
\bibitem [{\citenamefont {White}(1992)}]{White1992}%
  \BibitemOpen
  \bibfield  {author} {\bibinfo {author} {\bibfnamefont {S.~R.}\ \bibnamefont
  {White}},\ }\href {\doibase 10.1103/physrevlett.69.2863} {\bibfield
  {journal} {\bibinfo  {journal} {Physical Review Letters}\ }\textbf {\bibinfo
  {volume} {69}},\ \bibinfo {pages} {2863} (\bibinfo {year}
  {1992})}\BibitemShut {NoStop}%
\bibitem [{\citenamefont {White}\ and\ \citenamefont
  {Martin}(1999)}]{White1999}%
  \BibitemOpen
  \bibfield  {author} {\bibinfo {author} {\bibfnamefont {S.~R.}\ \bibnamefont
  {White}}\ and\ \bibinfo {author} {\bibfnamefont {R.~L.}\ \bibnamefont
  {Martin}},\ }\href {\doibase 10.1063/1.478295} {\bibfield  {journal}
  {\bibinfo  {journal} {The Journal of Chemical Physics}\ }\textbf {\bibinfo
  {volume} {110}},\ \bibinfo {pages} {4127} (\bibinfo {year}
  {1999})}\BibitemShut {NoStop}%
\bibitem [{\citenamefont {Chan}\ and\ \citenamefont
  {Head-Gordon}(2002)}]{Chan2002}%
  \BibitemOpen
  \bibfield  {author} {\bibinfo {author} {\bibfnamefont {G.~K.-L.}\
  \bibnamefont {Chan}}\ and\ \bibinfo {author} {\bibfnamefont {M.}~\bibnamefont
  {Head-Gordon}},\ }\href {\doibase 10.1063/1.1449459} {\bibfield  {journal}
  {\bibinfo  {journal} {The Journal of Chemical Physics}\ }\textbf {\bibinfo
  {volume} {116}},\ \bibinfo {pages} {4462} (\bibinfo {year}
  {2002})}\BibitemShut {NoStop}%
\bibitem [{\citenamefont {Legeza}, \citenamefont {R\"{o}der},\ and\
  \citenamefont {Hess}(2003)}]{Legeza2003a}%
  \BibitemOpen
  \bibfield  {author} {\bibinfo {author} {\bibfnamefont {{\"O}.}~\bibnamefont
  {Legeza}}, \bibinfo {author} {\bibfnamefont {J.}~\bibnamefont {R\"{o}der}}, \
  and\ \bibinfo {author} {\bibfnamefont {B.~A.}\ \bibnamefont {Hess}},\ }\href
  {\doibase 10.1103/physrevb.67.125114} {\bibfield  {journal} {\bibinfo
  {journal} {Physical Review B}\ }\textbf {\bibinfo {volume} {67}} (\bibinfo
  {year} {2003}),\ 10.1103/physrevb.67.125114}\BibitemShut {NoStop}%
\bibitem [{\citenamefont {Schollw\"{o}ck}(2011)}]{Schollwck2011}%
  \BibitemOpen
  \bibfield  {author} {\bibinfo {author} {\bibfnamefont {U.}~\bibnamefont
  {Schollw\"{o}ck}},\ }\href {\doibase 10.1016/j.aop.2010.09.012} {\bibfield
  {journal} {\bibinfo  {journal} {Annals of Physics}\ }\textbf {\bibinfo
  {volume} {326}},\ \bibinfo {pages} {96} (\bibinfo {year} {2011})}\BibitemShut
  {NoStop}%
\bibitem [{\citenamefont {Olivares-Amaya}\ \emph {et~al.}(2015)\citenamefont
  {Olivares-Amaya}, \citenamefont {Hu}, \citenamefont {Nakatani}, \citenamefont
  {Sharma}, \citenamefont {Yang},\ and\ \citenamefont
  {Chan}}]{OlivaresAmaya2015}%
  \BibitemOpen
  \bibfield  {author} {\bibinfo {author} {\bibfnamefont {R.}~\bibnamefont
  {Olivares-Amaya}}, \bibinfo {author} {\bibfnamefont {W.}~\bibnamefont {Hu}},
  \bibinfo {author} {\bibfnamefont {N.}~\bibnamefont {Nakatani}}, \bibinfo
  {author} {\bibfnamefont {S.}~\bibnamefont {Sharma}}, \bibinfo {author}
  {\bibfnamefont {J.}~\bibnamefont {Yang}}, \ and\ \bibinfo {author}
  {\bibfnamefont {G.~K.-L.}\ \bibnamefont {Chan}},\ }\href {\doibase
  10.1063/1.4905329} {\bibfield  {journal} {\bibinfo  {journal} {The Journal of
  Chemical Physics}\ }\textbf {\bibinfo {volume} {142}},\ \bibinfo {pages}
  {034102} (\bibinfo {year} {2015})}\BibitemShut {NoStop}%
\bibitem [{\citenamefont {Kurashige}\ and\ \citenamefont
  {Yanai}(2011)}]{Kurashige2011}%
  \BibitemOpen
  \bibfield  {author} {\bibinfo {author} {\bibfnamefont {Y.}~\bibnamefont
  {Kurashige}}\ and\ \bibinfo {author} {\bibfnamefont {T.}~\bibnamefont
  {Yanai}},\ }\href {\doibase 10.1063/1.3629454} {\bibfield  {journal}
  {\bibinfo  {journal} {The Journal of Chemical Physics}\ }\textbf {\bibinfo
  {volume} {135}},\ \bibinfo {pages} {094104} (\bibinfo {year}
  {2011})}\BibitemShut {NoStop}%
\bibitem [{\citenamefont {Saitow}, \citenamefont {Kurashige},\ and\
  \citenamefont {Yanai}(2013)}]{Saitow2013}%
  \BibitemOpen
  \bibfield  {author} {\bibinfo {author} {\bibfnamefont {M.}~\bibnamefont
  {Saitow}}, \bibinfo {author} {\bibfnamefont {Y.}~\bibnamefont {Kurashige}}, \
  and\ \bibinfo {author} {\bibfnamefont {T.}~\bibnamefont {Yanai}},\ }\href
  {\doibase 10.1063/1.4816627} {\bibfield  {journal} {\bibinfo  {journal} {The
  Journal of Chemical Physics}\ }\textbf {\bibinfo {volume} {139}},\ \bibinfo
  {pages} {044118} (\bibinfo {year} {2013})}\BibitemShut {NoStop}%
\bibitem [{\citenamefont {Wouters}\ \emph {et~al.}(2013)\citenamefont
  {Wouters}, \citenamefont {Nakatani}, \citenamefont {Neck},\ and\
  \citenamefont {Chan}}]{Wouters2013}%
  \BibitemOpen
  \bibfield  {author} {\bibinfo {author} {\bibfnamefont {S.}~\bibnamefont
  {Wouters}}, \bibinfo {author} {\bibfnamefont {N.}~\bibnamefont {Nakatani}},
  \bibinfo {author} {\bibfnamefont {D.~V.}\ \bibnamefont {Neck}}, \ and\
  \bibinfo {author} {\bibfnamefont {G.~K.-L.}\ \bibnamefont {Chan}},\ }\href
  {\doibase 10.1103/physrevb.88.075122} {\bibfield  {journal} {\bibinfo
  {journal} {Physical Review B}\ }\textbf {\bibinfo {volume} {88}} (\bibinfo
  {year} {2013}),\ 10.1103/physrevb.88.075122}\BibitemShut {NoStop}%
\bibitem [{\citenamefont {Yanai}\ and\ \citenamefont {Chan}(2006)}]{Yanai2006}%
  \BibitemOpen
  \bibfield  {author} {\bibinfo {author} {\bibfnamefont {T.}~\bibnamefont
  {Yanai}}\ and\ \bibinfo {author} {\bibfnamefont {G.~K.-L.}\ \bibnamefont
  {Chan}},\ }\href {\doibase 10.1063/1.2196410} {\bibfield  {journal} {\bibinfo
   {journal} {The Journal of Chemical Physics}\ }\textbf {\bibinfo {volume}
  {124}},\ \bibinfo {pages} {194106} (\bibinfo {year} {2006})}\BibitemShut
  {NoStop}%
\bibitem [{\citenamefont {Ren}, \citenamefont {Yi},\ and\ \citenamefont
  {Shuai}(2016)}]{Ren2016}%
  \BibitemOpen
  \bibfield  {author} {\bibinfo {author} {\bibfnamefont {J.}~\bibnamefont
  {Ren}}, \bibinfo {author} {\bibfnamefont {Y.}~\bibnamefont {Yi}}, \ and\
  \bibinfo {author} {\bibfnamefont {Z.}~\bibnamefont {Shuai}},\ }\href
  {\doibase 10.1021/acs.jctc.6b00696} {\bibfield  {journal} {\bibinfo
  {journal} {Journal of Chemical Theory and Computation}\ }\textbf {\bibinfo
  {volume} {12}},\ \bibinfo {pages} {4871} (\bibinfo {year}
  {2016})}\BibitemShut {NoStop}%
\bibitem [{\citenamefont {Veis}\ \emph {et~al.}(2016)\citenamefont {Veis},
  \citenamefont {Antal{\'{\i}}k}, \citenamefont {Brabec}, \citenamefont
  {Neese}, \citenamefont {Legeza},\ and\ \citenamefont {Pittner}}]{Veis2016}%
  \BibitemOpen
  \bibfield  {author} {\bibinfo {author} {\bibfnamefont {L.}~\bibnamefont
  {Veis}}, \bibinfo {author} {\bibfnamefont {A.}~\bibnamefont
  {Antal{\'{\i}}k}}, \bibinfo {author} {\bibfnamefont {J.}~\bibnamefont
  {Brabec}}, \bibinfo {author} {\bibfnamefont {F.}~\bibnamefont {Neese}},
  \bibinfo {author} {\bibfnamefont {{\"O}.}~\bibnamefont {Legeza}}, \ and\
  \bibinfo {author} {\bibfnamefont {J.}~\bibnamefont {Pittner}},\ }\href
  {\doibase 10.1021/acs.jpclett.6b01908} {\bibfield  {journal} {\bibinfo
  {journal} {The Journal of Physical Chemistry Letters}\ }\textbf {\bibinfo
  {volume} {7}},\ \bibinfo {pages} {4072} (\bibinfo {year} {2016})}\BibitemShut
  {NoStop}%
\bibitem [{\citenamefont {Kinoshita}, \citenamefont {Hino},\ and\ \citenamefont
  {Bartlett}(2005)}]{Kinoshita2005}%
  \BibitemOpen
  \bibfield  {author} {\bibinfo {author} {\bibfnamefont {T.}~\bibnamefont
  {Kinoshita}}, \bibinfo {author} {\bibfnamefont {O.}~\bibnamefont {Hino}}, \
  and\ \bibinfo {author} {\bibfnamefont {R.~J.}\ \bibnamefont {Bartlett}},\
  }\href {\doibase 10.1063/1.2000251} {\bibfield  {journal} {\bibinfo
  {journal} {The Journal of Chemical Physics}\ }\textbf {\bibinfo {volume}
  {123}},\ \bibinfo {pages} {074106} (\bibinfo {year} {2005})}\BibitemShut
  {NoStop}%
\bibitem [{\citenamefont {Hino}\ \emph {et~al.}(2006)\citenamefont {Hino},
  \citenamefont {Kinoshita}, \citenamefont {Chan},\ and\ \citenamefont
  {Bartlett}}]{Hino2006}%
  \BibitemOpen
  \bibfield  {author} {\bibinfo {author} {\bibfnamefont {O.}~\bibnamefont
  {Hino}}, \bibinfo {author} {\bibfnamefont {T.}~\bibnamefont {Kinoshita}},
  \bibinfo {author} {\bibfnamefont {G.~K.-L.}\ \bibnamefont {Chan}}, \ and\
  \bibinfo {author} {\bibfnamefont {R.~J.}\ \bibnamefont {Bartlett}},\ }\href
  {\doibase 10.1063/1.2180775} {\bibfield  {journal} {\bibinfo  {journal} {The
  Journal of Chemical Physics}\ }\textbf {\bibinfo {volume} {124}},\ \bibinfo
  {pages} {114311} (\bibinfo {year} {2006})}\BibitemShut {NoStop}%
\bibitem [{\citenamefont {Knecht}, \citenamefont {Legeza},\ and\ \citenamefont
  {Reiher}(2014)}]{Knecht2014}%
  \BibitemOpen
  \bibfield  {author} {\bibinfo {author} {\bibfnamefont {S.}~\bibnamefont
  {Knecht}}, \bibinfo {author} {\bibfnamefont {{\"O}.}~\bibnamefont {Legeza}},
  \ and\ \bibinfo {author} {\bibfnamefont {M.}~\bibnamefont {Reiher}},\ }\href
  {\doibase 10.1063/1.4862495} {\bibfield  {journal} {\bibinfo  {journal} {The
  Journal of Chemical Physics}\ }\textbf {\bibinfo {volume} {140}},\ \bibinfo
  {pages} {041101} (\bibinfo {year} {2014})}\BibitemShut {NoStop}%
\bibitem [{DIR()}]{DIRAC18}%
  \BibitemOpen
  \href@noop {} {}\bibinfo {note} {{DIRAC}, a relativistic ab initio electronic
  structure program (2018), T.~Saue et al.
  \url{http://www.diracprogram.org}}\BibitemShut {NoStop}%
\bibitem [{\citenamefont {Saue}\ and\ \citenamefont
  {Jensen}(1999)}]{saue-jensen-1999}%
  \BibitemOpen
  \bibfield  {author} {\bibinfo {author} {\bibfnamefont {T.}~\bibnamefont
  {Saue}}\ and\ \bibinfo {author} {\bibfnamefont {H.~J.~A.}\ \bibnamefont
  {Jensen}},\ }\href {\doibase 10.1063/1.479958} {\bibfield  {journal}
  {\bibinfo  {journal} {The Journal of Chemical Physics}\ }\textbf {\bibinfo
  {volume} {111}},\ \bibinfo {pages} {6211} (\bibinfo {year}
  {1999})}\BibitemShut {NoStop}%
\bibitem [{\citenamefont {Dyall}\ and\ \citenamefont
  {F{\ae}gri~Jr}(2007)}]{dyall-faegri}%
  \BibitemOpen
  \bibfield  {author} {\bibinfo {author} {\bibfnamefont {K.~G.}\ \bibnamefont
  {Dyall}}\ and\ \bibinfo {author} {\bibfnamefont {K.}~\bibnamefont
  {F{\ae}gri~Jr}},\ }\href@noop {} {\emph {\bibinfo {title} {Introduction to
  relativistic quantum chemistry}}}\ (\bibinfo  {publisher} {Oxford University
  Press},\ \bibinfo {year} {2007})\BibitemShut {NoStop}%
\bibitem [{\citenamefont {Thyssen}(2001)}]{Thyssen_phd}%
  \BibitemOpen
  \bibfield  {author} {\bibinfo {author} {\bibfnamefont {J.}~\bibnamefont
  {Thyssen}},\ }\emph {\bibinfo {title} {Development and Applications of
  Methods for Correlated Relativistic Calculations of Molecular Properties}},\
  \href@noop {} {Ph.D. thesis},\ \bibinfo  {school} {Univeristy of Southern
  Denmark} (\bibinfo {year} {2001})\BibitemShut {NoStop}%
\bibitem [{\citenamefont {Li}\ and\ \citenamefont
  {Paldus}(1997)}]{paldus-externalcorr}%
  \BibitemOpen
  \bibfield  {author} {\bibinfo {author} {\bibfnamefont {X.}~\bibnamefont
  {Li}}\ and\ \bibinfo {author} {\bibfnamefont {J.}~\bibnamefont {Paldus}},\
  }\href@noop {} {\bibfield  {journal} {\bibinfo  {journal} {J. Comp. Phys.}\
  }\textbf {\bibinfo {volume} {107}},\ \bibinfo {pages} {6257} (\bibinfo {year}
  {1997})}\BibitemShut {NoStop}%
\bibitem [{\citenamefont {Lyakh}, \citenamefont {Lotrich},\ and\ \citenamefont
  {Bartlett}(2011)}]{cyclobut-tailored-2011}%
  \BibitemOpen
  \bibfield  {author} {\bibinfo {author} {\bibfnamefont {D.~I.}\ \bibnamefont
  {Lyakh}}, \bibinfo {author} {\bibfnamefont {V.~F.}\ \bibnamefont {Lotrich}},
  \ and\ \bibinfo {author} {\bibfnamefont {R.~J.}\ \bibnamefont {Bartlett}},\
  }\href@noop {} {\bibfield  {journal} {\bibinfo  {journal} {Chem. Phys.
  Lett.}\ }\textbf {\bibinfo {volume} {501}},\ \bibinfo {pages} {166} (\bibinfo
  {year} {2011})}\BibitemShut {NoStop}%
\bibitem [{\citenamefont {Melnichuk}\ and\ \citenamefont
  {Bartlett}(2012)}]{melnichuk-2012}%
  \BibitemOpen
  \bibfield  {author} {\bibinfo {author} {\bibfnamefont {A.}~\bibnamefont
  {Melnichuk}}\ and\ \bibinfo {author} {\bibfnamefont {R.~J.}\ \bibnamefont
  {Bartlett}},\ }\href@noop {} {\bibfield  {journal} {\bibinfo  {journal} {J.
  Comp. Phys.}\ }\textbf {\bibinfo {volume} {137}},\ \bibinfo {pages} {214103}
  (\bibinfo {year} {2012})}\BibitemShut {NoStop}%
\bibitem [{\citenamefont {Melnichuk}\ and\ \citenamefont
  {Bartlett}(2014)}]{melnichuk-2014}%
  \BibitemOpen
  \bibfield  {author} {\bibinfo {author} {\bibfnamefont {A.}~\bibnamefont
  {Melnichuk}}\ and\ \bibinfo {author} {\bibfnamefont {R.~J.}\ \bibnamefont
  {Bartlett}},\ }\href@noop {} {\bibfield  {journal} {\bibinfo  {journal} {J.
  Comp. Phys.}\ }\textbf {\bibinfo {volume} {140}},\ \bibinfo {pages} {064113}
  (\bibinfo {year} {2014})}\BibitemShut {NoStop}%
\bibitem [{\citenamefont {Piecuch}, \citenamefont {Oliphant},\ and\
  \citenamefont {Adamowicz}(1993)}]{Piecuch1993}%
  \BibitemOpen
  \bibfield  {author} {\bibinfo {author} {\bibfnamefont {P.}~\bibnamefont
  {Piecuch}}, \bibinfo {author} {\bibfnamefont {N.}~\bibnamefont {Oliphant}}, \
  and\ \bibinfo {author} {\bibfnamefont {L.}~\bibnamefont {Adamowicz}},\ }\href
  {\doibase 10.1063/1.466179} {\bibfield  {journal} {\bibinfo  {journal} {The
  Journal of Chemical Physics}\ }\textbf {\bibinfo {volume} {99}},\ \bibinfo
  {pages} {1875} (\bibinfo {year} {1993})}\BibitemShut {NoStop}%
\bibitem [{\citenamefont {Piecuch}\ and\ \citenamefont
  {Adamowicz}(1994)}]{semi3}%
  \BibitemOpen
  \bibfield  {author} {\bibinfo {author} {\bibfnamefont {P.}~\bibnamefont
  {Piecuch}}\ and\ \bibinfo {author} {\bibfnamefont {L.}~\bibnamefont
  {Adamowicz}},\ }\href {\doibase 10.1063/1.467143} {\bibfield  {journal}
  {\bibinfo  {journal} {The Journal of Chemical Physics}\ }\textbf {\bibinfo
  {volume} {100}},\ \bibinfo {pages} {5792} (\bibinfo {year} {1994})},\ \Eprint
  {http://arxiv.org/abs/https://doi.org/10.1063/1.467143}
  {https://doi.org/10.1063/1.467143} \BibitemShut {NoStop}%
\bibitem [{\citenamefont {Kowalski}(2018)}]{Kowalski2018}%
  \BibitemOpen
  \bibfield  {author} {\bibinfo {author} {\bibfnamefont {K.}~\bibnamefont
  {Kowalski}},\ }\href {\doibase 10.1063/1.5010693} {\bibfield  {journal}
  {\bibinfo  {journal} {The Journal of Chemical Physics}\ }\textbf {\bibinfo
  {volume} {148}},\ \bibinfo {pages} {094104} (\bibinfo {year}
  {2018})}\BibitemShut {NoStop}%
\bibitem [{\citenamefont {Faulstich}\ \emph
  {et~al.}(2019{\natexlab{a}})\citenamefont {Faulstich}, \citenamefont
  {Laestadius}, \citenamefont {\"{O}rs Legeza}, \citenamefont {Schneider},\
  and\ \citenamefont {Kvaal}}]{Faulstich2019a}%
  \BibitemOpen
  \bibfield  {author} {\bibinfo {author} {\bibfnamefont {F.~M.}\ \bibnamefont
  {Faulstich}}, \bibinfo {author} {\bibfnamefont {A.}~\bibnamefont
  {Laestadius}}, \bibinfo {author} {\bibnamefont {\"{O}rs Legeza}}, \bibinfo
  {author} {\bibfnamefont {R.}~\bibnamefont {Schneider}}, \ and\ \bibinfo
  {author} {\bibfnamefont {S.}~\bibnamefont {Kvaal}},\ }\href {\doibase
  10.1137/18m1171436} {\bibfield  {journal} {\bibinfo  {journal} {{SIAM}
  Journal on Numerical Analysis}\ }\textbf {\bibinfo {volume} {57}},\ \bibinfo
  {pages} {2579} (\bibinfo {year} {2019}{\natexlab{a}})}\BibitemShut {NoStop}%
\bibitem [{\citenamefont {Veis}\ \emph {et~al.}(2017)\citenamefont {Veis},
  \citenamefont {Antal{\'{\i}}k}, \citenamefont {Brabec}, \citenamefont
  {Neese}, \citenamefont {Legeza},\ and\ \citenamefont
  {Pittner}}]{Veis2016err}%
  \BibitemOpen
  \bibfield  {author} {\bibinfo {author} {\bibfnamefont {L.}~\bibnamefont
  {Veis}}, \bibinfo {author} {\bibfnamefont {A.}~\bibnamefont
  {Antal{\'{\i}}k}}, \bibinfo {author} {\bibfnamefont {J.}~\bibnamefont
  {Brabec}}, \bibinfo {author} {\bibfnamefont {F.}~\bibnamefont {Neese}},
  \bibinfo {author} {\bibfnamefont {{\"O}.}~\bibnamefont {Legeza}}, \ and\
  \bibinfo {author} {\bibfnamefont {J.}~\bibnamefont {Pittner}},\ }\href@noop
  {} {\bibfield  {journal} {\bibinfo  {journal} {The Journal of Physical
  Chemistry Letters}\ }\textbf {\bibinfo {volume} {8}},\ \bibinfo {pages} {291}
  (\bibinfo {year} {2017})}\BibitemShut {NoStop}%
\bibitem [{\citenamefont {Legeza}\ and\ \citenamefont
  {S{\'{o}}lyom}(2003)}]{Legeza2003}%
  \BibitemOpen
  \bibfield  {author} {\bibinfo {author} {\bibfnamefont {{\"O}.}~\bibnamefont
  {Legeza}}\ and\ \bibinfo {author} {\bibfnamefont {J.}~\bibnamefont
  {S{\'{o}}lyom}},\ }\href {\doibase 10.1103/physrevb.68.195116} {\bibfield
  {journal} {\bibinfo  {journal} {Physical Review B}\ }\textbf {\bibinfo
  {volume} {68}} (\bibinfo {year} {2003}),\
  10.1103/physrevb.68.195116}\BibitemShut {NoStop}%
\bibitem [{\citenamefont {Schollw\"ock}(2005)}]{schollwock_2005}%
  \BibitemOpen
  \bibfield  {author} {\bibinfo {author} {\bibfnamefont {U.}~\bibnamefont
  {Schollw\"ock}},\ }\href@noop {} {\bibfield  {journal} {\bibinfo  {journal}
  {Rev. Mod. Phys.}\ }\textbf {\bibinfo {volume} {77}},\ \bibinfo {pages} {259}
  (\bibinfo {year} {2005})}\BibitemShut {NoStop}%
\bibitem [{\citenamefont {Legeza}\ \emph {et~al.}(2008)\citenamefont {Legeza},
  \citenamefont {Noack}, \citenamefont {S\'olyom},\ and\ \citenamefont
  {Tincani}}]{Legeza-2008}%
  \BibitemOpen
  \bibfield  {author} {\bibinfo {author} {\bibfnamefont {{\"O}.}~\bibnamefont
  {Legeza}}, \bibinfo {author} {\bibfnamefont {R.}~\bibnamefont {Noack}},
  \bibinfo {author} {\bibfnamefont {J.}~\bibnamefont {S\'olyom}}, \ and\
  \bibinfo {author} {\bibfnamefont {L.}~\bibnamefont {Tincani}},\ }in\
  \href@noop {} {\emph {\bibinfo {booktitle} {Computational Many-Particle
  Physics}}},\ \bibinfo {series} {Lecture Notes in Physics}, Vol.\ \bibinfo
  {volume} {739},\ \bibinfo {editor} {edited by\ \bibinfo {editor}
  {\bibfnamefont {H.}~\bibnamefont {Fehske}}, \bibinfo {editor} {\bibfnamefont
  {R.}~\bibnamefont {Schneider}}, \ and\ \bibinfo {editor} {\bibfnamefont
  {A.}~\bibnamefont {Weisse}}}\ (\bibinfo  {publisher} {Springer Berlin
  Heidelberg},\ \bibinfo {year} {2008})\ pp.\ \bibinfo {pages}
  {653--664}\BibitemShut {NoStop}%
\bibitem [{\citenamefont {Marti}\ and\ \citenamefont
  {Reiher}(2010)}]{marti_2010}%
  \BibitemOpen
  \bibfield  {author} {\bibinfo {author} {\bibfnamefont {K.~H.}\ \bibnamefont
  {Marti}}\ and\ \bibinfo {author} {\bibfnamefont {M.}~\bibnamefont {Reiher}},\
  }\href@noop {} {\bibfield  {journal} {\bibinfo  {journal} {Z. Phys. Chem.}\
  }\textbf {\bibinfo {volume} {224}},\ \bibinfo {pages} {583} (\bibinfo {year}
  {2010})}\BibitemShut {NoStop}%
\bibitem [{\citenamefont {Chan}\ and\ \citenamefont
  {Sharma}(2011)}]{chan_review}%
  \BibitemOpen
  \bibfield  {author} {\bibinfo {author} {\bibfnamefont {G.~K.-L.}\
  \bibnamefont {Chan}}\ and\ \bibinfo {author} {\bibfnamefont {S.}~\bibnamefont
  {Sharma}},\ }\href@noop {} {\bibfield  {journal} {\bibinfo  {journal} {Ann.
  Rev. Phys. Chem.}\ }\textbf {\bibinfo {volume} {62}},\ \bibinfo {pages} {465}
  (\bibinfo {year} {2011})}\BibitemShut {NoStop}%
\bibitem [{\citenamefont {Wouters}\ and\ \citenamefont
  {Van~Neck}(2014)}]{wouters_review}%
  \BibitemOpen
  \bibfield  {author} {\bibinfo {author} {\bibfnamefont {S.}~\bibnamefont
  {Wouters}}\ and\ \bibinfo {author} {\bibfnamefont {D.}~\bibnamefont
  {Van~Neck}},\ }\href@noop {} {\bibfield  {journal} {\bibinfo  {journal} {Eur.
  Phys. J. D}\ }\textbf {\bibinfo {volume} {68}},\ \bibinfo {eid} {272}
  (\bibinfo {year} {2014})}\BibitemShut {NoStop}%
\bibitem [{\citenamefont {Szalay}\ \emph {et~al.}(2015)\citenamefont {Szalay},
  \citenamefont {Pfeffer}, \citenamefont {Murg}, \citenamefont {Barcza},
  \citenamefont {Verstraete}, \citenamefont {Schneider},\ and\ \citenamefont
  {Legeza}}]{Szalay2015}%
  \BibitemOpen
  \bibfield  {author} {\bibinfo {author} {\bibfnamefont {S.}~\bibnamefont
  {Szalay}}, \bibinfo {author} {\bibfnamefont {M.}~\bibnamefont {Pfeffer}},
  \bibinfo {author} {\bibfnamefont {V.}~\bibnamefont {Murg}}, \bibinfo {author}
  {\bibfnamefont {G.}~\bibnamefont {Barcza}}, \bibinfo {author} {\bibfnamefont
  {F.}~\bibnamefont {Verstraete}}, \bibinfo {author} {\bibfnamefont
  {R.}~\bibnamefont {Schneider}}, \ and\ \bibinfo {author} {\bibfnamefont
  {{\"O}.}~\bibnamefont {Legeza}},\ }\href {\doibase 10.1002/qua.24898}
  {\bibfield  {journal} {\bibinfo  {journal} {International Journal of Quantum
  Chemistry}\ }\textbf {\bibinfo {volume} {115}},\ \bibinfo {pages} {1342}
  (\bibinfo {year} {2015})}\BibitemShut {NoStop}%
\bibitem [{\citenamefont {Yanai}\ \emph {et~al.}(2015)\citenamefont {Yanai},
  \citenamefont {Kurashige}, \citenamefont {Mizukami}, \citenamefont
  {Chalupsk\'y}, \citenamefont {Lan},\ and\ \citenamefont
  {Saitow}}]{yanai_review}%
  \BibitemOpen
  \bibfield  {author} {\bibinfo {author} {\bibfnamefont {T.}~\bibnamefont
  {Yanai}}, \bibinfo {author} {\bibfnamefont {Y.}~\bibnamefont {Kurashige}},
  \bibinfo {author} {\bibfnamefont {W.}~\bibnamefont {Mizukami}}, \bibinfo
  {author} {\bibfnamefont {J.}~\bibnamefont {Chalupsk\'y}}, \bibinfo {author}
  {\bibfnamefont {T.~N.}\ \bibnamefont {Lan}}, \ and\ \bibinfo {author}
  {\bibfnamefont {M.}~\bibnamefont {Saitow}},\ }\href@noop {} {\bibfield
  {journal} {\bibinfo  {journal} {Int. J. Quant. Chem.}\ }\textbf {\bibinfo
  {volume} {115}},\ \bibinfo {pages} {283} (\bibinfo {year}
  {2015})}\BibitemShut {NoStop}%
\bibitem [{\citenamefont {Moritz}\ and\ \citenamefont
  {Reiher}(2007)}]{moritz_2007}%
  \BibitemOpen
  \bibfield  {author} {\bibinfo {author} {\bibfnamefont {G.}~\bibnamefont
  {Moritz}}\ and\ \bibinfo {author} {\bibfnamefont {M.}~\bibnamefont
  {Reiher}},\ }\href@noop {} {\bibfield  {journal} {\bibinfo  {journal} {J.
  Chem. Phys.}\ }\textbf {\bibinfo {volume} {126}},\ \bibinfo {pages} {244109}
  (\bibinfo {year} {2007})}\BibitemShut {NoStop}%
\bibitem [{\citenamefont {Boguslawski}, \citenamefont {Marti},\ and\
  \citenamefont {Reiher}(2011)}]{boguslawski_2011}%
  \BibitemOpen
  \bibfield  {author} {\bibinfo {author} {\bibfnamefont {K.}~\bibnamefont
  {Boguslawski}}, \bibinfo {author} {\bibfnamefont {K.~H.}\ \bibnamefont
  {Marti}}, \ and\ \bibinfo {author} {\bibfnamefont {M.}~\bibnamefont
  {Reiher}},\ }\href@noop {} {\bibfield  {journal} {\bibinfo  {journal} {J.
  Chem. Phys.}\ }\textbf {\bibinfo {volume} {134}},\ \bibinfo {pages} {224101}
  (\bibinfo {year} {2011})}\BibitemShut {NoStop}%
\bibitem [{\citenamefont {Zgid}\ and\ \citenamefont
  {Nooijen}(2008)}]{Zgid-2008b}%
  \BibitemOpen
  \bibfield  {author} {\bibinfo {author} {\bibfnamefont {D.}~\bibnamefont
  {Zgid}}\ and\ \bibinfo {author} {\bibfnamefont {M.}~\bibnamefont {Nooijen}},\
  }\href {\doibase http://dx.doi.org/10.1063/1.2883980} {\bibfield  {journal}
  {\bibinfo  {journal} {The Journal of Chemical Physics}\ }\textbf {\bibinfo
  {volume} {128}},\ \bibinfo {eid} {144115} (\bibinfo {year}
  {2008})}\BibitemShut {NoStop}%
\bibitem [{\citenamefont {Battaglia}, \citenamefont {Keller},\ and\
  \citenamefont {Knecht}(2018)}]{Battaglia2018}%
  \BibitemOpen
  \bibfield  {author} {\bibinfo {author} {\bibfnamefont {S.}~\bibnamefont
  {Battaglia}}, \bibinfo {author} {\bibfnamefont {S.}~\bibnamefont {Keller}}, \
  and\ \bibinfo {author} {\bibfnamefont {S.}~\bibnamefont {Knecht}},\ }\href
  {\doibase 10.1021/acs.jctc.7b01065} {\bibfield  {journal} {\bibinfo
  {journal} {Journal of Chemical Theory and Computation}\ }\textbf {\bibinfo
  {volume} {14}},\ \bibinfo {pages} {2353} (\bibinfo {year}
  {2018})}\BibitemShut {NoStop}%
\bibitem [{\citenamefont {Legeza}, \citenamefont {Veis},\ and\ \citenamefont
  {Mosoni}()}]{budapest_qcdmrg}%
  \BibitemOpen
  \bibfield  {author} {\bibinfo {author} {\bibfnamefont {{\"O}.}~\bibnamefont
  {Legeza}}, \bibinfo {author} {\bibfnamefont {L.}~\bibnamefont {Veis}}, \ and\
  \bibinfo {author} {\bibfnamefont {T.}~\bibnamefont {Mosoni}},\ }\href@noop {}
  {\enquote {\bibinfo {title} {{QC-DMRG-Budapest, a program for quantum
  chemical DMRG calculations}},}\ }\BibitemShut {NoStop}%
\bibitem [{\citenamefont {Faulstich}\ \emph
  {et~al.}(2019{\natexlab{b}})\citenamefont {Faulstich}, \citenamefont
  {M{\'{a}}t{\'{e}}}, \citenamefont {Laestadius}, \citenamefont {Csirik},
  \citenamefont {Veis}, \citenamefont {Antalik}, \citenamefont {Brabec},
  \citenamefont {Schneider}, \citenamefont {Pittner}, \citenamefont {Kvaal},\
  and\ \citenamefont {Legeza}}]{Faulstich2019b}%
  \BibitemOpen
  \bibfield  {author} {\bibinfo {author} {\bibfnamefont {F.~M.}\ \bibnamefont
  {Faulstich}}, \bibinfo {author} {\bibfnamefont {M.}~\bibnamefont
  {M{\'{a}}t{\'{e}}}}, \bibinfo {author} {\bibfnamefont {A.}~\bibnamefont
  {Laestadius}}, \bibinfo {author} {\bibfnamefont {M.~A.}\ \bibnamefont
  {Csirik}}, \bibinfo {author} {\bibfnamefont {L.}~\bibnamefont {Veis}},
  \bibinfo {author} {\bibfnamefont {A.}~\bibnamefont {Antalik}}, \bibinfo
  {author} {\bibfnamefont {J.}~\bibnamefont {Brabec}}, \bibinfo {author}
  {\bibfnamefont {R.}~\bibnamefont {Schneider}}, \bibinfo {author}
  {\bibfnamefont {J.}~\bibnamefont {Pittner}}, \bibinfo {author} {\bibfnamefont
  {S.}~\bibnamefont {Kvaal}}, \ and\ \bibinfo {author} {\bibfnamefont
  {{\"O}.}~\bibnamefont {Legeza}},\ }\href {\doibase 10.1021/acs.jctc.8b00960}
  {\bibfield  {journal} {\bibinfo  {journal} {Journal of Chemical Theory and
  Computation}\ }\textbf {\bibinfo {volume} {15}},\ \bibinfo {pages} {2206}
  (\bibinfo {year} {2019}{\natexlab{b}})}\BibitemShut {NoStop}%
\bibitem [{\citenamefont {Titov}\ \emph {et~al.}(2001)\citenamefont {Titov},
  \citenamefont {Mosyagin}, \citenamefont {Alekseyev},\ and\ \citenamefont
  {Buenker}}]{Titov2001}%
  \BibitemOpen
  \bibfield  {author} {\bibinfo {author} {\bibfnamefont {A.~V.}\ \bibnamefont
  {Titov}}, \bibinfo {author} {\bibfnamefont {N.~S.}\ \bibnamefont {Mosyagin}},
  \bibinfo {author} {\bibfnamefont {A.~B.}\ \bibnamefont {Alekseyev}}, \ and\
  \bibinfo {author} {\bibfnamefont {R.~J.}\ \bibnamefont {Buenker}},\ }\href
  {\doibase 10.1002/1097-461x(2001)81:6<409::aid-qua1010>3.0.co;2-0} {\bibfield
   {journal} {\bibinfo  {journal} {International Journal of Quantum Chemistry}\
  }\textbf {\bibinfo {volume} {81}},\ \bibinfo {pages} {409} (\bibinfo {year}
  {2001})}\BibitemShut {NoStop}%
\bibitem [{\citenamefont {Hensel}, \citenamefont {Hughes},\ and\ \citenamefont
  {Brown}(1995)}]{Hensel1995}%
  \BibitemOpen
  \bibfield  {author} {\bibinfo {author} {\bibfnamefont {K.~D.}\ \bibnamefont
  {Hensel}}, \bibinfo {author} {\bibfnamefont {R.~A.}\ \bibnamefont {Hughes}},
  \ and\ \bibinfo {author} {\bibfnamefont {J.~M.}\ \bibnamefont {Brown}},\
  }\href {\doibase 10.1039/ft9959102999} {\bibfield  {journal} {\bibinfo
  {journal} {J. Chem. Soc., Faraday Trans.}\ }\textbf {\bibinfo {volume}
  {91}},\ \bibinfo {pages} {2999} (\bibinfo {year} {1995})}\BibitemShut
  {NoStop}%
\bibitem [{\citenamefont {Balasubramanian}(1989)}]{Balasubramanian1989}%
  \BibitemOpen
  \bibfield  {author} {\bibinfo {author} {\bibfnamefont {K.}~\bibnamefont
  {Balasubramanian}},\ }\href {\doibase 10.1021/cr00098a008} {\bibfield
  {journal} {\bibinfo  {journal} {Chemical Reviews}\ }\textbf {\bibinfo
  {volume} {89}},\ \bibinfo {pages} {1801} (\bibinfo {year}
  {1989})}\BibitemShut {NoStop}%
\bibitem [{\citenamefont {Beutel}\ \emph {et~al.}(1996)\citenamefont {Beutel},
  \citenamefont {Setzer}, \citenamefont {Shestakov},\ and\ \citenamefont
  {Fink}}]{Beutel1996}%
  \BibitemOpen
  \bibfield  {author} {\bibinfo {author} {\bibfnamefont {M.}~\bibnamefont
  {Beutel}}, \bibinfo {author} {\bibfnamefont {K.}~\bibnamefont {Setzer}},
  \bibinfo {author} {\bibfnamefont {O.}~\bibnamefont {Shestakov}}, \ and\
  \bibinfo {author} {\bibfnamefont {E.}~\bibnamefont {Fink}},\ }\href {\doibase
  10.1006/jmsp.1996.0186} {\bibfield  {journal} {\bibinfo  {journal} {Journal
  of Molecular Spectroscopy}\ }\textbf {\bibinfo {volume} {179}},\ \bibinfo
  {pages} {79} (\bibinfo {year} {1996})}\BibitemShut {NoStop}%
\bibitem [{\citenamefont {Urban}, \citenamefont {Essig},\ and\ \citenamefont
  {Jones}(1993)}]{Urban1993}%
  \BibitemOpen
  \bibfield  {author} {\bibinfo {author} {\bibfnamefont {R.-D.}\ \bibnamefont
  {Urban}}, \bibinfo {author} {\bibfnamefont {K.}~\bibnamefont {Essig}}, \ and\
  \bibinfo {author} {\bibfnamefont {H.}~\bibnamefont {Jones}},\ }\href
  {\doibase 10.1063/1.465327} {\bibfield  {journal} {\bibinfo  {journal} {The
  Journal of Chemical Physics}\ }\textbf {\bibinfo {volume} {99}},\ \bibinfo
  {pages} {1591} (\bibinfo {year} {1993})}\BibitemShut {NoStop}%
\end{thebibliography}%

\end{document}